\documentclass[10pt,conference]{IEEEtran}
\IEEEoverridecommandlockouts

\makeatletter
\def\endthebibliography{%
	\def\@noitemerr{\@latex@warning{Empty `thebibliography' environment}}%
	\endlist
}
\makeatother

\usepackage{cite}
\usepackage{makecell}
\usepackage{amsmath,amssymb,amsfonts}
\usepackage{algorithmic}
\usepackage{url}
\usepackage{graphicx}
\usepackage{textcomp}
\usepackage{enumerate}
\usepackage{authblk}
\usepackage[tikz]{ocgx2}
\usepackage{pgfplotstable}
\pgfplotsset{width=5cm, height=10cm}
\usepackage{xcolor}
\usepackage{float}
\usepackage{colortbl}
\usepackage{hyperref}
\usepackage[capitalise]{cleveref}
\usepackage{siunitx}

\def\BibTeX{{\rm B\kern-.05em{\sc i\kern-.025em b}\kern-.08em
		T\kern-.1667em\lower.7ex\hbox{E}\kern-.125emX}}

\begin{document}
	
	\title{Memory Error Detection in Security Testing}
	\author[]{Nasif Imtiaz}
	\author[]{Laurie Williams}
	\affil[]{Department of Computer Science, North Carolina State University}
	\affil[]{simtiaz@ncsu.edu, lawilli3@ncsu.edu}
	\maketitle
	
	\begin{abstract}
	We study 10 C/C++ projects that have been using a static analysis security testing tool. We analyze the historical scan reports generated by the tool and study how frequently memory-related alerts appeared. We also studied the subsequent developer action on those alerts. We also look at the CVEs published for these projects within study timeline and investigate how many of them are memory related. Moreover, for one of this project, Linux, we investigate if the involved flaws in the CVE were identified by the studied security tool when they were first introduced in the code. \\
	We found memory related alerts to be frequently detected during static analysis security testing. However, based on how actively the project developers are monitoring the tool alerts, these errors can take years to get fixed. For the ten studied projects, we found a median lifespan of 77 days before memory alerts get fixed. We also find that around 40\% of the published CVEs for the studied C/C++ projects are related to memory. These memory CVEs have higher CVSS severity ratings and likelihood of having an exploit script public than non-memory CVEs. We also found only 2.5\% Linux CVEs were possibly detected during static analysis security testing.
	\end{abstract}
	
	
	\section{Introduction}
	
	Software development is an error-prone activity. Code written and reviewed by skilled developers may end up with various types of bugs some of which can be security-critical, i.e. vulnerability. Developers use security testing tools (e.g. static analysis, fuzzing) in order to identify the potential vulnerabilities. However, given the non-perfect nature of these tools and the cost of fixing the alerts they generate, developer response to such alerts may vary. Further, even after extensive security testing, undetected vulnerabilities may still remain in the code that get discovered years after it was introduced.

Memory-related bugs, such as buffer overflows and uninitialized reads, are an important class of security vulnerabilities typically more prevalent in non memory-safe languages such as C, C++. Security testing techniques such as static analysis can detect a wide range of potential memory issues in the code. Tools performing such analysis present their results as alerts to the developers. However, it is yet to be studied how frequently do memory-related alerts get identified by security tools and how developers respond to these alerts. Further, vulnerabilities that leak through all security testing, may get discovered years later. If reported to the National Vulnerability Database (NVD), the discovered vulnerabilities are tracked in a central database with a unique identifier called CVE. However, it is yet to be studied if these vulnerabilities (CVEs) appeared in the code because security testing techniques failed to identify them or the testing could have had indeed identified the flaw but the developers did not take a corrective action in time.

In this paper, we study 10 C/C++ projects that have been using a static analysis security testing tool. We analyze the historical scan reports generated by the tool and study how frequently memory-related alerts appeared. We also studied the subsequent developer action on those alerts. We also look at the CVEs published for these projects within study timeline and investigate how many of them are memory related. Moreover, for one of this project, Linux, we investigate if the involved flaws in the CVE were identified by the studied security tool when they were first introduced in the code. We state our research questions as:

\begin{itemize}
    \item \textbf{RQ1:} How frequently do memory-related alerts get identified by a static analysis security testing tool? How do developers respond to these alerts?
    \item \textbf{RQ2:} What portion of publicly known vulnerabilities, CVEs, are memory related? How many of these CVEs were identified by a static analysis security testing tool when the involved flaws were first introduced in the code?
\end{itemize}

\section{Dataset}
We primarily work with two datasets:
\begin{figure*}
    \centering
    \includegraphics[scale=.1]{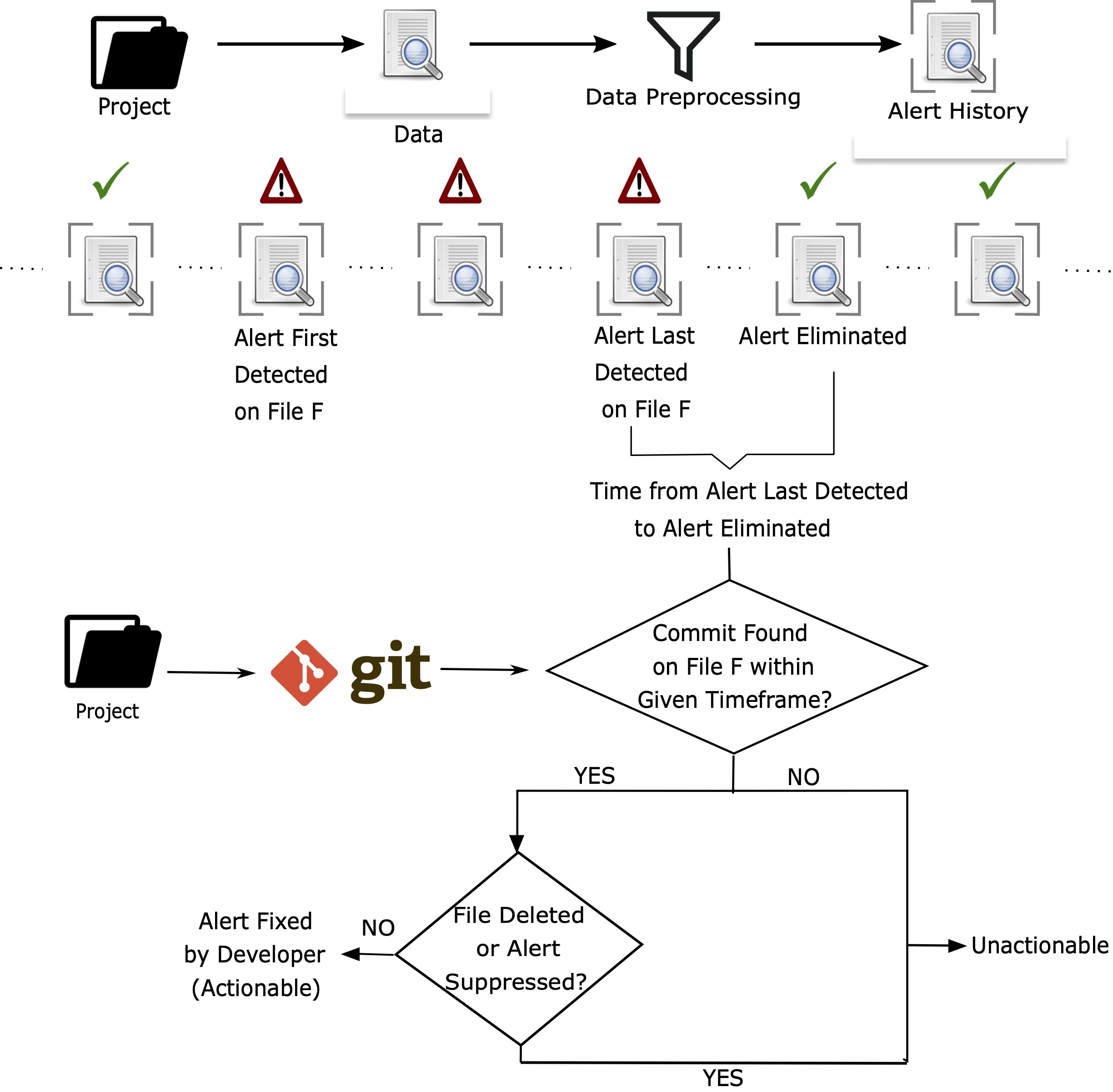}
    \caption{Determining actionability of static analysis alerts}
    \label{fig:method}
\end{figure*}
\begin{table*}[]
    \centering
    \caption{Studied projects}
    \label{tab:projectinfo}
    \begin{tabular}{llllrlrl}
 Project               & Scan Reports   & Start Date   & End Date   &   Scan Interval (days) & Total                      Alerts   &   Triaged Alerts (\%) & Lines of Code   \\
\hline
 Linux                 & 648            & 2012-05-17   & 2020-06-15 &                      3 & 19,514                              &                13.66 & 13,959,859      \\
 Firefox               & 662            & 2006-02-22   & 2018-10-27 &                      2 & 12,944                              &                36.33 & 8,223,984       \\
 LibreOffice           & 414            & 2012-10-13   & 2020-06-26 &                      4 & 11,982                              &                55.42 & 6,130,847       \\
 Samba                 & 770            & 2006-02-23   & 2020-06-17 &                      3 & 4,507                               &                52.58 & 2,941,352       \\
 VTK                   & 74             & 2015-12-21   & 2017-08-12 &                      7 & 2,522                               &                 2.50 & 2,354,516       \\
 OpenCV                & 553            & 2012-12-12   & 2020-06-30 &                      2 & 2,443                               &                 9.41 & 1,264,056       \\
 Kodi                  & 489            & 2012-08-28   & 2020-05-16 &                      3 & 2,393                               &                42.42 & 743,623         \\
 !CHAOS Control System & 170            & 2015-01-14   & 2019-03-19 &                      1 & 1,980                               &                12.37 & 880,411         \\
 Chromium EC           & 1,713          & 2016-01-15   & 2020-07-06 &                      0 & 964                                 &                28.11 & 61,040          \\
 Thunderbird           & 438            & 2006-04-12   & 2020-02-29 &                      1 & 885                                 &                27.23 & 640,424         \\
\hline
\end{tabular}
\end{table*}
\begin{enumerate}
    \item A proprietary dataset that consists of historical scan reports for 10 C/C++ projects by a static analysis security testing tool, we refer to this dataset as 'scan history' dataset;
    \item The CVE dataset from National Vulnerability Database~\cite{nvdcve}.
Alerts in scan history dataset and CVEs in NVD database have corresponding CWE mapping (Common Weakness Enumeration, CWE, is a list of software and hardware weaknesses~\cite{cwe}). Two human reviwers independently classified the CWE identifiers as memory vs. non-memory to separate out the memory-related issues in our dataset.
\end{enumerate}

The following subsections explain the data sources, and subsequent CWE classification.

\subsection{Scan History Dataset}
We study historical scan reports by an industry-leading static analysis security testing tool for ten C/C++ projects. For each project, we have a range of scan reports between a certain time period with short intervals for the latest code version on the corresponding scan date. We confirmed that developers of respective projects monitor these scan reports by looking at the alert triage rate (how many alerts were triaged by developers).

For each scan report, the alerts are tagged as either i) Fixed (the security tool stopped detecting the alert in the code, we will refer to this as eliminated), ii) Dismissed (When developers have explicitly marked an alert as False Positive or Intentional), or iii) New (the security tool still detects the alert in the current scan, we refer to these as alive alerts). For the eliminated alerts, to distinguish between developer fix and other modes of elimination (e.g. alert suppression, file deletion/renaming), we track any commit made on the affected files for an alert to see if there were any valid code change made when the alert was marked as fixed in a subsequent scan. The alerts that were fixed through code change made by developers, we refer to them as actionable alerts. We also calculated the lifespan of each alert (the time span the security tool kept detecting an alert), and the fix complexity for the actionable alerts (change in files, lines of code, and logical blocks of code in the commit that fixed the alert). Figure \ref{fig:method} explains the workflow. The details of projects used in this paper are listed in Table \ref{tab:projectinfo}.

\subsection{CVE dataset}
For the ten projects from scan history dataset, we intended to get the published CVEs since the first scan report of the dataset. We searched the CVEs from NVD API through the Common Product Enumeration (CPE) identifier for each project. However, for Chromium EC, VTK, and !CHAOS Control System, we could not find a matching CPE identifier. We pulled the CVEs for rest of the seven projects.

Further, we augmented the CVE dataset with patch commits for a CVE whenever available (Patch commit is the exact commit that were merged to the product code to fix a vulnerability). We looked for the patch commits for each CVE through the reference links provided in the CVE data and respective project repositories. For Firefox, LibreOffice, and Thunderbird, we could not find the commits while Kodi had only 3 CVEs in our dataset (all are non memory-related). While it is possible to collect patch commits in the case of Samba, and OpenCV, they also have low number of CVEs published. Therefore, we only focus on Linux when answering parts of RQ2.

We also looked for publicly available exploit scripts for the CVEs in our dataset. We use a open source tool, cve\_searchsploit~\footnote{\url{https://github.com/andreafioraldi/cve_searchsploit}}, in order to find the available exploit scripts. The tool pulls its data from publicly curated 'Exploit Database'~\footnote{https://www.exploit-db.com/}.
\begin{table}[]
    \centering
    \caption{Memory related alerts}
    \label{tab:f1}
    \begin{tabular}{lll}
\hline
 Project               & Total Alerts   & Memory Alerts   \\
\hline
Linux                 & 19,514         & 7,903 (40.5\%)   \\
 Firefox               & 12,944         & 5,448 (42.09\%)  \\
 LibreOffice           & 11,982         & 3,975 (33.17\%)  \\
 Samba                 & 4,507          & 1,700 (37.72\%)  \\
 VTK                   & 2,522          & 1,166 (46.23\%)  \\
 OpenCV                & 2,443          & 710 (29.06\%)    \\
 Kodi                  & 2,393          & 682 (28.5\%)     \\
 !CHAOS Control System & 1,980          & 282 (14.24\%)    \\
 Chromium EC           & 964            & 306 (31.74\%)    \\
 Thunderbird           & 885            & 357 (40.34\%)    \\
\hline
\end{tabular}
\end{table}
\subsection{CWE classification for memory vs. non-memory}
In order to focus on memory-related errors, we classified the alerts in scan history dataset and the CVEs as memory and non-memory through their corresponding CWE identifier. CWE~\cite{cwe} is a list of common weakness types in software and serves as a common measuring stick for security tools. Each CWE identifier has a name, description, and various details on how to detect, mitigate, or exploit software weaknesses. Two human reviewers, one graduate student and one industry professional, independently reviewed the CWE ids and classified them as memory or non-memory. We defined CWEs as memory-related when a weakness can make software to access memory in a way that was unintended by the developer. These weaknesses include access errors, use of uninitialized variables, and memory leaks which can all lead to accessing system memory in an unintended way.

There are 161 distinct CWE identifiers covering both of our data set. Two reviewers independently classified them as memory vs. non-memory with a Kohen's Kappa agreement rate of 0.79, which is interpreted as substantial agreement [4]. On the CWE identifiers that were classified differently, the reviewers had a discussion and came to a negotiated agreement for a final classification. We classified 43 CWE ids as memory (26.7\%) and the rest as non-memory.
\begin{table*}[]
    \centering
    \caption{Actionability of memory alerts}
    \label{tab:f2}
    \begin{tabular}{lrlll}
\hline
 Project               &   Total Memory Alerts & Eliminated   & Actionable & Triaged Bug by Devs.   \\
\hline
 Linux                 &                  7903 & 73.06\%              & 43.54\%              & 3.11\%                         \\
 Firefox               &                  5448 & 77.22\%              & 49.72\%              & 12.43\%                        \\
 LibreOffice           &                  3975 & 80.63\%              & 73.46\%              & 51.35\%                        \\
 Samba                 &                  1700 & 77.88\%              & 39.88\%              & 3.82\%                         \\
 VTK                   &                  1166 & 12.78\%              & 7.2\%                & 0.43\%                         \\
 OpenCV                &                   710 & 89.15\%              & 32.54\%              & 0.28\%                         \\
 Kodi                  &                   682 & 68.62\%              & 54.11\%              & 25.07\%                        \\
 Thunderbird           &                   357 & 63.31\%              & 24.37\%              & 12.61\%                        \\
 Chromium EC           &                   306 & 93.46\%              & 29.74\%              & 17.65\%                        \\
 !CHAOS Control System &                   282 & 80.14\%              & 39.01\%              & 1.42\%                         \\                        \\
\hline
\end{tabular}
\end{table*}
\begin{table*}[]
    \centering
    \caption{Unactionable memory alerts}
    \label{tab:f3}
\begin{tabular}{llllll}
\hline
 Project        & Total memory alerts   & Marked intentional   & Marked false positive  & Alive at last scan   & \makecell{Eliminated through \\undetermined ways } \\
\hline
 Linux       & 7,903          & 1.64\%                & 4.02\%                   & 21.08\%         & 28.22\%                       \\
 Firefox     & 5,448          & 3.82\%                & 9.69\%                   & 8.87\%          & 25.11\%                       \\
 LibreOffice & 3,975          & 0.78\%                & 18.39\%                  & 0.08\%          & 7.02\%                        \\
 Samba       & 1,700          & 0.82\%                & 2.88\%                   & 18.12\%         & 35.53\%                       \\
 VTK         & 1,166          & 0.09\%                & 0.17\%                   & 86.96\%         & 4.8\%                         \\
 OpenCV      & 710            & 0.56\%                & 1.97\%                   & 8.31\%          & 41.55\%                       \\
 Kodi        & 682            & 10.56\%               & 9.38\%                   & 11.44\%         & 9.09\%                        \\
 Thunderbird & 357            & 0.28\%                & 0.28\%                   & 31.37\%         & 38.94\%                       \\
 Chromium EC & 306            & 0.33\%                & 2.61\%                   & 3.59\%          & 63.73\%                       \\
\hline
\end{tabular}
\end{table*}
\section{Findings: How frequently do memory-related alerts get identified by a static analysis security testing tool? How do developers respond to these alerts?}

We investigate the scan history dataset to answer the following questions:

\subsection{To what extent do memory related alerts appear when C/C++ projects are scanned by a static analysis tool?}
While the rate of alerts that are memory related vary among projects, approximately one-third of static analysis alerts were found to be memory related (median rate of 35.4\% among ten projects). Table \ref{tab:f1} lists the rate of memory related alerts for each project.

\subsection{What percent of memory related alerts are actionable (i.e. acted on by developers)? What percent of memory related alerts were marked as false positives by the developers?}
We then looked at how many memory related alerts were eliminated (elimination does not necessarily mean developer fix), how many were acted on by developers (fixed through valid code change), and how many were explicitly triaged as bug by the respective project developers. We find 39.9\% (median among ten projects) memory alerts to be acted on by the developers while 12.4\% of them are explicitly marked as bug. We also compared the elimination rate, actionable, and bug rate between memory and non-memory alerts through Mann-Whitney U test for the ten projects. However, we do not find any statistical significant difference between memory and non-memory alerts. Table \ref{tab:f2} lists details of actionable memory alerts for each studied project.

We then looked at the unactionable memory alerts. Developers in our dataset had the ability to mark an alert as False Positive or Intentional which gives us direct feedback about the alerts from the respective project developers. We see that memory alerts are rarely to be intentional coding (median rate of 0.8\%) and are significantly less than non-memory alerts (5.7\%). However, we do not find any difference in being false positive or alive between memory and non-memory related alerts. Table \ref{tab:f3} lists details of unactionable memory alerts for each studied project.

\subsection{What is the lifespan of memory related alerts on the codebase?}
We define lifespan of an alert as the difference in days between when an alert was first detected and when it was last detected. Below tables present lifespan in days for actionable alerts and then specifically the alerts that were triaged as bug by the project developers with a breakdown between memory vs. non-memory. The median lifespan among ten projects for actionable memory alerts are 77 days and 98.5 days when the alert was triaged as bug. The difference in lifespan for memory and non-memory alerts were statistically insignificant. Table \ref{tab:l1} and \ref{tab:l2} lists lifespan of actionable alerts for each project.

\subsection{What is the fix complexity of memory related alerts on the codebase?}
In our prior work~\cite{imtiaz2019developers}, we found fix of static analysis are low in complexity; a median of 4 lines of code change in the affected file. We find similar results for memory alerts in this paper. When reporting median, for memory alerts: we find 2.5 files are changed with 6.6 lines of code change and 2 blocks of logical change (change in consecutive lines). Within the affected file, there is 3.5 lines of code change and 1 block of logical change.

\subsection{What is the prevalence and developer response of alerts in terms of CWE identifiers?}
Here, we present a breakdown of our prior analyses in term of CWE-identifier. Table \ref{tab:cwe} presents information for top 10 CWE-identifiers prevalent in scan history dataset with 'Null Pointer Dereference' being the most common alert.
\begin{table*}[]
    \centering
    \caption{Lifespan of actionable alerts}
    \label{tab:l1}
    \begin{tabular}{lrrr}
\hline
 Project               &   \makecell{Lifespan of all\\ Actionable alerts (days)} &   \makecell{Lifespan of\\ Memory alerts(days)} &   \makecell{Lifespan of \\Non-memory alerts (days)} \\
\hline
Thunderbird           &                                      915.5 &                               1591 &                                    490 \\
 OpenCV                &                                      345.5 &                                342 &                                    380 \\
 Linux                 &                                      258   &                                257 &                                    258 \\
 !CHAOS Control System &                                      136.5 &                                 48 &                                    203 \\
 Firefox               &                                      124   &                                 44 &                                    154 \\
 ovirt-engine          &                                      102   &                                 77 &                                    119 \\
 Chromium EC           &                                       93   &                                247 &                                     70 \\
 Samba                 &                                       70   &                                143 &                                     37 \\
 Kodi                  &                                       62   &                                  6 &                                     82 \\
 VTK                   &                                       31   &                                 24 &                                     42 \\
 LibreOffice           &                                       18   &                                 16 &                                     18 \\
\hline
\end{tabular}
\end{table*}
\begin{table*}[]
    \centering
    \caption{Lifespan of alerts marked as bug}
    \label{tab:l2}
    \begin{tabular}{lrrr}
\hline
 Project      &   \makecell{Lifespan of alerts\\ marked as bug (days)} &   \makecell{Lifespan of\\ Memory alerts (days)} &   \makecell{Lifespan of \\Non-memory alerts (days)}\\
\hline
Thunderbird  &                                      1385 &                             1779   &                                  970   \\
 Samba        &                                       665 &                              665   &                                  911   \\
 Chromium EC  &                                       377 &                              429   &                                  367   \\
 Linux        &                                       190 &                              134   &                                  252   \\
 Firefox      &                                        64 &                               63   &                                   71.5 \\
 ovirt-engine &                                        46 &                               39.5 &                                   49.5 \\
 VTK          &                                        24 &                               24   &                                   30   \\
 LibreOffice  &                                        15 &                               21   &                                    6   \\
 OpenCV       &                                         5 &                              850.5 &                                    5   \\
 Kodi         &                                         2 &                                2   &                                    2   \\

\hline
\end{tabular}
\end{table*}

\begin{table*}[]
    \centering
    \caption{Top CWE categories of alerts}
    \label{tab:cwe}
    \begin{tabular}{rp{5cm}lllrrl}
\hline
   CEW-Id & CWE-name                                                                & No. of alerts   & Eliminated   & Actionable   &   lifespan (days) &   \makecell{Triaged\\ Bug by Devs.} & \makecell{Triaged\\ False Positive}   \\
\hline
            476 & NULL Pointer Dereference                                                & 8,689 (14.45\%)  & 81.86\%       & 58.42\%       &              44   &                  21.99 & 4.12\%                    \\
      404 & Improper Resource Shutdown or Release                                   & 4,250 (7.07\%)   & 66.92\%       & 43.41\%       &             122   &                  14.33 & 18.99\%                   \\
      119 & Improper Restriction of Operations within the Bounds of a Memory Buffer & 2,125 (3.53\%)   & 67.06\%       & 33.22\%       &             278   &                   4.42 & 12.89\%                   \\
      457 & Use of Uninitialized Variable                                           & 1,862 (3.1\%)    & 74.01\%       & 53.49\%       &             135   &                  15.52 & 6.12\%                    \\
      190 & Integer Overflow or Wraparound                                          & 1,003 (1.67\%)   & 54.24\%       & 35.39\%       &             315   &                   7.88 & 7.48\%                    \\
      125 & Out-of-bounds Read                                                      & 939 (1.56\%)     & 67.41\%       & 42.81\%       &             244   &                   8.2  & 11.82\%                   \\
      416 & Use After Free                                                          & 877 (1.46\%)     & 73.43\%       & 36.26\%       &              71   &                   7.41 & 19.27\%                   \\
      120 & Buffer Copy without Checking Size of Input ('Classic Buffer Overflow')  & 751 (1.25\%)     & 78.7\%        & 15.98\%       &             515   &                   2.13 & 3.6\%                     \\
      170 & Improper Null Termination                                               & 378 (0.63\%)     & 68.78\%       & 56.88\%       &             379.5 &                   7.94 & 9.26\%                    \\
      590 & Free of Memory not on the Heap                                          & 363 (0.6\%)      & 50.96\%       & 30.03\%       &             322   &                   0    & 6.89\%                    \\
\hline
\end{tabular}
\end{table*}

\section{Findings: What portion of publicly known vulnerabilities, CVEs, are memory related? How many of these CVEs were identified by a static analysis security testing tool when the involved flaws were first introduced in the code?}
Common Vulnerabilities and Exposures (CVEs) are a reference method for publicly known software vulnerabilities[2]. While softwares typically go through security testing before being released, vulnerabilities may still remain in the code either due to testing techniques not having a perfect recall or inadequate developer corrective action. In this paper, we investigate how many CVEs could have been identified by a security tool when the bug was first introduced in order to estimate the contribution of testing inefficacy vs. developer inaction.

In the following subsections, we first investigate how many CVEs were memory related for the projects in scan history dataset. We then choose Linux as a case study to determine how many of its CVEs were identified by the studied static analysis security testing tool when first introduced.

\subsection{What percent of CVEs are memory related?}
We pulled CVEs for seven projects from the scan history dataset that were published after the first scan and before the last scan report in our dataset. We further classified CVEs as memory vs. non-memory based on their corresponding CWE ids. Table \ref{tab:cve1} shows a breakdown of CVEs for these projects.

As a median for the seven projects, we find that 40\% of the CVEs are memory related. We further look at the severity ratings, in the form of CVSS2 and CVSS3 scores for these CVEs in Table \ref{tab:sev}.  We find that median CVSS2 score for memory alerts for all seven projects is 7.35 which is significantly greater than non-memory alerts (5.0). Similarly for CVSS3 rating, memory alerts median rating is 9.3, significantly larger than non-memory alerts (6.6)

\begin{table}[]
    \centering
    \caption{Memory related CVEs}
    \label{tab:cve1}
    \begin{tabular}{lll}
\hline
 Project     & \makecell{No. of CVEs \\(within study timeline)}   & memory-related CVEs   \\
\hline
 Linux       & 3,144                                 & 45.4\%                 \\
 Firefox     & 1,731                                 & 36.0\%                 \\
 Thunderbird & 940                                   & 45.3\%                 \\
 Samba       & 138                                   & 29.7\%                 \\
 LibreOffice & 40                                    & 40.0\%                 \\
 OpenCV      & 33                                    & 75.8\%                 \\
 Kodi        & 3                                     & 0.0\%                  \\
\hline
\end{tabular}
\end{table}
\begin{table*}[]
    \centering
    \caption{Severity scores of Memory and Non-memory CVEs}
    \label{tab:sev}
    \begin{tabular}{lrrrrrr}
\hline
 Project     &   Memory CVEs &   Median CVSS2 score &   Median CVSS3 score &   Non-Memory CVEs &   Median CVSS2 score &   Median CVSS3 score  \\
\hline
 Linux       &     1427 &           7.2  &            7.8 &        1150 &              4.9  &               6.7 \\
 Firefox     &      624 &           7.5  &            9.8 &         737 &              5    &               6.1 \\
 Thunderbird &      426 &           7.55 &            9.8 &         308 &              5    &               6.5 \\
 Samba       &       41 &           5    &            6.5 &          84 &              5    &               6.5 \\
 OpenCV      &       25 &           6.8  &            8.8 &           6 &              4.65 &               6.7 \\
 LibreOffice &       16 &           7.5  &            9.8 &          20 &              6.8  &               7.8 \\
\hline
\end{tabular}
    
\end{table*}

\subsection{What percent of memory related CVEs have exploit scripts available?}
If all vulnerabilities can be exploited in the wild to attack the involved system is debatable. However, for many CVEs, there are exploit scripts publicly available. We find that for 396 CVEs among the seven projects, there is an exploit script available out of which 234 (59 \%) are memory related. Table \ref{tab:exploit} present a CWE-breakdown for memory CVEs with an exploit.

\begin{table*}[]
    \centering
     \caption{CVEs with exploit script available}
    \label{tab:exploit}
    \begin{tabular}{rll}
\hline
   CWE-id & CWE-name                                                                & portion among memory CVEs   \\
\hline
      119 & Improper Restriction of Operations within the Bounds of a Memory Buffer & 145 (62.0\%)                 \\
      416 & Use After Free                                                          & 41 (17.5\%)                  \\
      787 & Out-of-bounds Write                                                     & 13 (5.6\%)                   \\
      190 & Integer Overflow or Wraparound                                          & 9 (3.8\%)                    \\
      125 & Out-of-bounds Read                                                      & 7 (3.0\%)                    \\
      824 & Access of Uninitialized Pointer                                         & 4 (1.7\%)                    \\
      476 & NULL Pointer Dereference                                                & 3 (1.3\%)                    \\
      122 & Heap-based Buffer Overflow                                              & 2 (0.9\%)                    \\
      191 & Integer Underflow (Wrap or Wraparound)                                  & 2 (0.9\%)                    \\
      415 & Double Free                                                             & 2 (0.9\%)                    \\
\hline
\end{tabular}
   
\end{table*}

\subsection{How many Linux CVEs were identified by a static analysis tool when the vulnerability was first introduced in the code?}
For this question, we focus on only Linux product. In the NVD data feed for Linux product, there can be link to the patch commit (the code change merged into codebase to fix a CVE) among the reference links. For 1,427 Linux memory related CVEs, we found patch commits for 592 CVEs. The rest of the CVEs were mostly in external product code (e.g. flash player) and not in Linux code itself.

We then analyzed the patch commit(s) for each CVE, to extract the names of the files that were changed in the fix with an assumption that the vulnerability related code are among these files. We then look at the time when these patch commit(s) were merged into the main codebase of Linux. If there were a memory alert in scan history dataset on the involved files of the CVE, the alert should get fixed when the patch commit is merged (alert last appears before patch commit is merged, and get fixed after the merge). If there is such memory alerts on the involved files that get fixed when a CVE get fixed, we assume that the studied security tool had an alert relevant to the certain CVE.

Only for 15 CVEs (2.5\%), we found there was a possible static analysis alert. We further looked at the 15 cases manually by looking at the CVE description and type of the alert raised by the static analysis tool. For 12 cases, the alert type somewhat matches CVE description, while we could not determine such matching for rest of the three cases. We further looked at the lifespan for these 15 alerts and found their median lifespan to be 567 days.

\section{Conclusion}

We found memory related alerts to be frequently detected during static analysis security testing. However, based on how actively the project developers are monitoring the tool alerts, these errors can take years to get fixed. For the ten studied projects, we found a median lifesspan of 77 days before memory alerts get fixed.

We also find that around 40\% of the published CVEs for the studied C/C++ projects are related to memory. These memory CVEs have higher CVSS severity ratings and likelihood of having an exploit script public than non-memory CVEs. We also found only 2.5\% Linux CVEs were possibly detected during static analysis security testing.

	\bibliographystyle{plain}
	\bibliography{bibliography.bib}

\end{document}